\documentstyle[11pt,cal97,psfig]{article}

\begin{document}

\title{The Wavelength Calibration of the WFC grism} 

\author{A. Pasquali, N. Pirzkal and J.R. Walsh}
\affil{ESO/ST-ECF, Karl-Schwarzschild-Strasse 2, D-85748 Garching
       bei M\"unchen, Germany}

\begin{abstract}
We present the wavelength solution derived for the G800l grism  
with the Wide Field Channel from the spectra of two Galactic Wolf-Rayet
stars, WR~45 and WR~96. The data were acquired in-orbit during the SMOV
tests and the early INTERIM calibration program. We have obtained an
average dispersion of 39.2 \AA/pix in the first order, 20.5 \AA/pix in
the second and -42.5 \AA/pix in the negative first order. We show that
the wavelength solution is strongly field-dependent, with an amplitude
of the variation of about 11$\%$ from the center of the WFC aperture to
the corners. The direction of the field-dependence is the diagonal from 
the image left top corner (amplifier A) to the bottom right corner
(amplifier D). These trends are observed for all grism orders. We also
describe the calibration files derived from the SMOV and INTERIM data which
are used by the ST-ECF slitless extraction code aXe.  
\end{abstract}

\keywords{grism spectroscopy, slitless spectroscopy}

\section{Introduction}
The Advanced Camera for Surveys (ACS) has been designed to perform 
low-resolution, slitless spectroscopy over a wide range of
wavelengths, from the Ly$\alpha$ line at $\lambda$ =
1216 \AA\ to $\sim$ 1 $\mu$m. One {\it optical} grism, one
{\it blue} prism and two {\it near-UV} prisms cover this 
range and are coupled with the Wide Field (WFC) and
High Resolution (HRC) Channels, the HRC and the Solar Blind
Channel (SBC) respectively.
\par\noindent
The WFC and the HRC make use of the same grism, G800L which works
between $\sim$ 5500 \AA\ and $\sim$ 1 $\mu$m. Its nominal dispersion
is $\sim$ 40 \AA/pix and $\sim$ 29 \AA/pix in first order for
the WFC and HRC respectively.
\par\noindent
The HRC also features a prism, PR200L, which covers the spectral range
between $\sim$ 2000 \AA\ and 5000 \AA, with a non linear dispersion
varying from 2.6 \AA/pix at $\lambda$ = 1600 \AA\ to 91 \AA/pix
at $\lambda$ = 3500 \AA\ and 515 \AA/pix at $\lambda$ = 5000 \AA.
\par\noindent
The SBC is equipped with two prisms, PR110L and PR130L which range 
from $\sim$ 1150 \AA\ and 1250 \AA\ to 2000 \AA\ with a resolving power of $\sim$ 80
and $\sim$ 100, respectively, at $\lambda$ = 1600 \AA. In particular,
PR130L does not include the geocoronal L$\alpha$ line for low background
measurements.

Pasquali et al. (2001b) showed that the high angular resolution of
the ACS may easily decrease the effective resolution of the grism,
since, when no slit is used, the grism nominal resolution is
convolved with the object size along the dispersion axis. The
extension of any grism spectrum along the cross-dispersed direction
is set by the size of the object which acts as an extraction
aperture. This is also an additional source of degradation when
the whole spectrum is summed along the cross-dispersion axis. 

The amplitude of these effects was investigated by simulating with
SLIM 1.0 (Pirzkal et al. 2001b) the spectrum of the Galactic
Planetary Nebula NGC 7009, and by increasing the linear size of 
the object as well as its orientation in the sky. The simulated grism
spectra indicated that line blending becomes severe when objects
are observed with a diameter larger than 2 pixels (0''.1) and
with a major axis at PA $>$ 45$^o$ with respect to the dispersion
axis (cf. Pasquali et al. 2001b).

These limits pose strong constraints on the selection of targets for
the in-orbit wavelength calibration of the ACS spectral elements.
Indeed, such calibrators have to be sorted by:
\begin{enumerate}
\item high brightness, to allow for short exposure times and 
time-series observations across the field of view;
\item a large number of emission lines in their spectra;
\item the absence of an extended nebula, which would otherwise
degrade the spectral resolution;
\item  negligible spectro-photometric variability, to be able
to identify emission lines at any observation epoch;
\item minimum field crowding, to avoid contamination by spectra
of nearby stars;
\item visibility, to allow repeated HST visits.
\end{enumerate}

The above set of requirements rules out Planetary Nebulae (PNe) as
possible wavelength calibrators, at least in the case of the
optical G800L grism. Indeed, PNe are resolved by HST up to 
the Large Magellanic Clouds and hence do not meet requirement $\#$ 3,
while PNe in M31 are compact enough but faint and therefore can
not fulfil requirements $\#$ 1, 6 and 5 as they also lie in crowded fields
(Pasquali et al. 2001a).

Wolf-Rayet stars (WRs) of spectral type WC have been shown to satisfy
all the requirements, at the expense of introducing a further constraint
which concerns the velocity of their stellar wind. Indeed, the wind
velocity in WRs can be as slow as 700 km~s$^{-1}$ and as fast as 3300
km~s$^{-1}$ (cf. van der Hucht 2001). A typical wind speed of 2000 km~s$^{-1}$
produces a line broadening of about 1.3 and 1.9 pixels in the grism first 
order with the WFC and the HRC, respectively. Therefore, to limit the
loss of resolution due to objects with broad emission lines, WR
stars should be selected with V$_{wind} \le$ 2000 km~s$^{-1}$ (Pasquali
et al. 2001a).
 
\section{The observational strategy}

We eventually selected two Galactic WR stars from the VII$^{th}$ Catalogue 
by van der Hucht (2001) which meet the listed criteria. Their basic properties,
coordinates, V magnitude and wind velocity are in Table 1.

\begin{table}
\caption{The WR stars selected for the wavelength calibration of the ACS grism.}
\begin{center}
\begin{tabular}{cccccc}
Star & Spectral & RA (2000) & DEC (2000) & V mag & Wind speed\\
     & type     &           &            &       & (km~s$^{-1}$)\\
\tableline
WR~45 & WC6 & 11:38:05.2 & -62:16:01 & 14.80 & 2100 \\
WR~96 & WC9 & 17:36:24.2 & -32:54:29 & 14.14 & 1100\\
\tableline
\end{tabular}
\end{center}
\end{table}

Both stars had been observed from the ground with the ESO/NTT EMMI spectrograph
with the purpose to acquire high-resolution spectra which would be later used as 
templates for the comparison with the ACS grism observations. The EMMI spectra
of WR~45 and WR~96 are plotted in Figure 1, where the dispersion is 1.26 \AA/pix.
  
\begin{figure}
\centerline{
\psfig{file=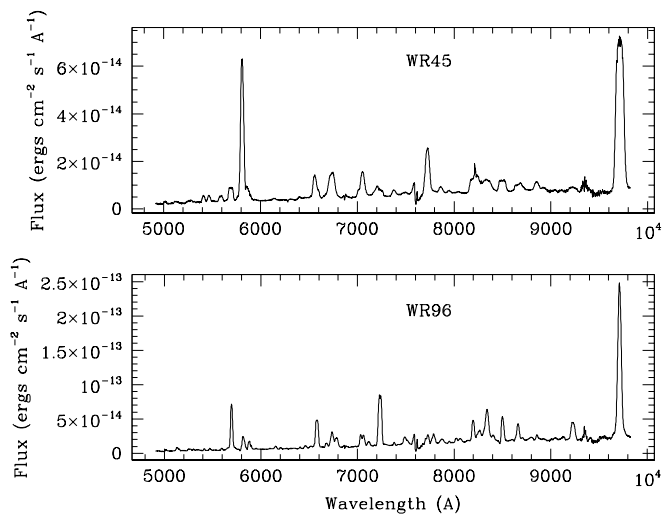,height=7cm}}
\caption{The spectra of WR~45 and WR~96 acquired with the ESO/NTT EMMI spectrograph
with a dispersion of 1.26 \AA/pix.}
\end{figure}

\subsection{Observations during the SMOV tests}

WR~45 was observed as part of the Servicing Mission Orbital Verification (SMOV) 
tests (ID 9029, PI Pasquali), at the end of April to early May 2002. Spectra
were taken at five different pointings across the field of view (f.o.v) of
the WFC: W1 close to the center of Chip 2, W3 and W5 close to amplifiers C and
D of Chip2 and W7 and W9 close to amplifiers A and B in Chip 1. These
pointings are shown in Figure 2.

\begin{figure}
\centerline{
\psfig{file=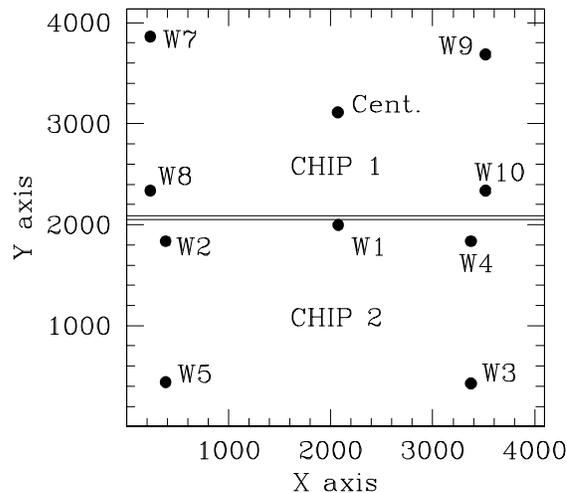}}
\caption{The pointings across the f.o.v of the WFC used during the SMOV and
INTERIM observations.}
\end{figure}

At each position, a pair of direct and grism images were acquired, and 
repeated two to four times, either in the same visit or in a subsequent one to
verify the stability of the filter wheel positioning. The
direct image, which provides the zero point of the grism dispersion correction,
was taken in the F625W and F775W filters, in order to check the target
position stability with wavelength. The adopted exposure times were 1 s 
for the direct imaging and 20 s for the grism.

\subsection{Observations during the INTERIM program}

WR~96 was observed during the INTERIM calibrations (ID 9568, PI Pasquali)
in June 2002. The observational strategy was similar to WR~45, but the number of individual
pointings was increased to 10 by adding to the SMOV positions the W2, W4, W8,
W10 pointings and the centre of Chip 1 (cf. Figure 2).

Monodimensional spectra of WR~45 and WR~96 were extracted from the raw, non drizzled 
images using the ST-ECF slitless spectra extraction code, aXe  
(Pirzkal et al. 2001a, http://www.stecf.org/software/aXe/index.html).
 
\section{The grism characteristics}

The extraction of slitless spectra relies on a number of parameters: 
\begin{enumerate}
\item the shift in
the X and Y coordinates between the position of the target in the direct image
and the position of the zeroth order in the grism image;
\item the tilt of spectra;
\item the separation in pixels along the dispersion axis of the {\it nth} grism
order from the zeroth;
\item the length in pixels along the dispersion axis of each grism order.
\end{enumerate}

While the shift allows a spectrum to be identified in the grism image given the
coordinates of the target in the direct image, the tilt enables it to be traced.
The separation and the length of the grism orders are used to set the
extraction aperture for each order spectrum.
Because of the severe geometric distortions in the WFC, these quantities are
expected to be field-dependent.

\subsection{The X- and Y-shifts}

The X- and Y-shifts could be measured for all the pointings, except 
W7 and W8 whose zeroth orders fall outside the physical boundaries of the
grism image. The remaining are listed in Table 2 in units of pixels;
they are the difference between the target position in the direct image
and the zeroth order coordinate in the grism image.
The values are averages among multiple measurements available
for each pointing. The standard deviation is typically 0.1 pixels.

\begin{table}
\caption{The X- and Y-shifts and the spectra tilt across the WFC aperture.}
\begin{center}
\begin{tabular}{cccccc}
Position & X-shift & Y-shift & Tilt \\
         & (in pixels) & (in pixels) & (in degrees)\\
\tableline
W1 & 113.02 & -3.14 & -1.91 \\
W2 & 120.02 & -3.56 & -1.95 \\
W3 & 102.02 & -1.85 & -1.43 \\
W4 & 106.32 & -2.88 & -1.85 \\
W5 & 115.24 & -2.77 & -1.53 \\
W7 &        &       & -2.52 \\
W8 &        &       & -2.11 \\
W9 & 112.83 & -4.57 & -2.32 \\
W10 & 108.15 & -3.39 & -1.95 \\
Chip1 center & 117.32 & -4.18 & -2.25 \\
\tableline
\end{tabular}
\end{center}
\end{table}

A decrease can be recognized for the Y-shift 
along the diagonal from amplifier A (W7 position) to D (W3 position).

\subsection{The tilt}

The tilt of the spectra was derived by fitting the (X,Y) coordinates of the emission
line peaks along the dispersion axis, measured from the negative third to the positive
third order (the negative orders at  smaller x pixels than the zeroth order x coordinate,
the positive ones at larger x pixels) . A first order polynomial was used to determine the slope of
the spectra with respect the X axis. Repeated measurements were averaged and
the standard deviation of the spectra tilt was determined to be typically of
0$^o$.02. The average tilt is shown in Table 2 as a function of position
in the WFC.

On average, the tilt of the grism spectra in the WFC is -1$^o$.98, and it
is field-dependent as it increases along the W7 - W3 diagonal by $\simeq$ 1$^o$.1. 

\subsection{The separation and length of the grism orders}

The distance in pixels of the grism orders from the zeroth order and their approximate
length were measured by counting the pixels along the X axis whose counts
are 3$\sigma$ above the background level. The mean length and FWHM of
the zeroth order are 23 and 4.4 pixels, respectively. The
average (across the f.o.v of the WFC) order separations and lengths are
listed in Table 3 in pixels.

\begin{table}
\caption{The separation from the zeroth order and the length in pixels of
the grism orders (from the SMOV data).}
\begin{center}
\begin{tabular}{ccccc}
Parameter & 1$^{st}$ ord. & 2$^{nd}$ ord. & -1$^{st}$ ord. & -2$^{nd}$ ord. \\
\tableline
Separation & 93 & 251 & -122 & -247\\
Length    & 156 & 125 & 102 & 111\\
\tableline
\end{tabular}
\end{center}
\end{table}

\section{The method of wavelength calibration}

The grism spectra were extracted in both units of pixels and wavelength
adopting for the latter the wavelength solutions derived from the ground
calibrations of the ACS. This allowed us to derive the mean FWHM in
\AA\ of the lines in each grism order. The NTT/EMMI template spectra
of WR~45 and WR~96 were then convolved by these mean FWHMs, their lines
reidentified and the line wavelengths re-measured. The 
position in pixels of the same lines was measured in the ACS grism spectra with respect
to the target X position in the direct image and tables of pixels
vs. wavelengths were built. Each table was then fitted with the routine
POLYFIT in IRAF and a wavelength solution was determined 
for each grism order. 

This procedure was applied to each grism spectrum in all positions 
across the f.o.v of the WFC.

\section{The wavelength solutions for the WFC and G800L grism}
In this proceeding we present the wavelength solutions (and
their field-dependence) computed for the grism first, second and
negative first orders which are the brightest. A field map of
the dispersion correction for the grism third,
negative second and negative third orders can be found in Pasquali
et al. (2003).

\subsection{The Grism first order}

An example of the grism first order as obtained in position W1 and
at the centre of Chip 1 is shown in Figure 3 for WR~45 
(left) and WR~96 (right).

\begin{figure}
\centerline{
\psfig{file=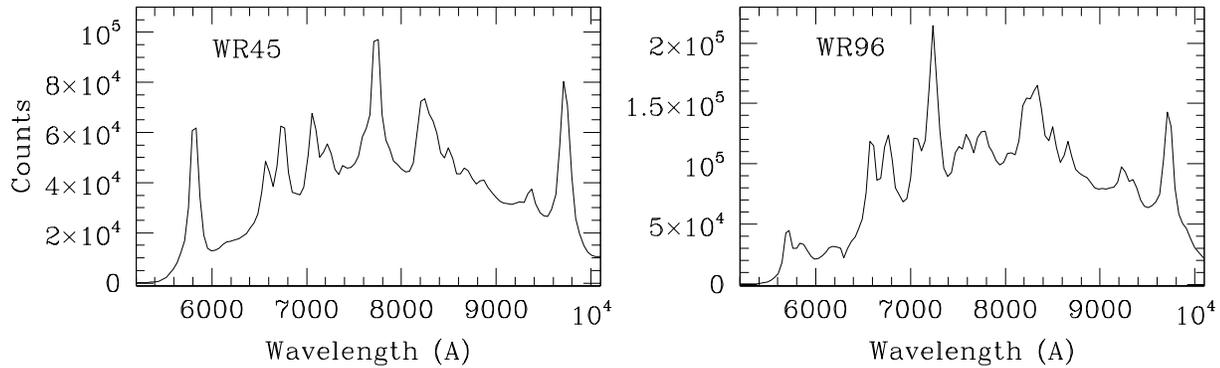}}
\caption{The grism first order spectra acquired for WR~45 at position W1 (left)
and WR~96 at the centre of Chip 1 (right).}
\end{figure}

The dispersion correction for the grism first order is reproduced by
a second order polynomial of the form: $\lambda$ = $\lambda_0 + \Delta\lambda_0$X $+
\Delta\lambda_1$X$^2$, where X is the distance along the dispersion axis from the
target X position in the direct image.

The wavelength solutions obtained for the ten pointings are reported in
Table 4. The tabulated values are averages of multiple measurements of the dispersion 
parameters derived for each pointing. The typical RMS associated with the fits
is 3 \AA/pix, while the typical error on $\lambda_0$ is 7 \AA. The uncertainty on the
first-oder term of the dispersion ($\Delta\lambda_0$) is 0.2 \AA/pix.

\begin{table}
\caption{The wavelength solutions obtained for the grism first order
as a function of positon across the f.o.v of the WFC.}
\begin{center}
\begin{tabular}{lccc}
Position & $\lambda_0$ & $\Delta\lambda_0$ & $\Delta\lambda_1$\\
         & (\AA)         & (\AA/pix)           & (\AA/pix$^2$) \\
\tableline
W1   & 4815.25 & 39.79 & 0.0099 \\
W2   & 4777.62 & 37.28 & 0.0098 \\ 
W3   & 4811.80 & 44.03 & 0.0096 \\
W4   & 4760.06 & 41.94 & 0.0108 \\
W5   & 4803.95 & 39.13 & 0.0097 \\
W7   & 4800.86 & 35.09 & 0.0068 \\
W8   & 4772.51 & 36.23 & 0.0098 \\
W9   & 4795.39 & 39.64 & 0.0095 \\
W10  & 4777.90 & 40.83 & 0.0130 \\
Chip 1 center& 4787.27 & 37.82 & 0.0112\\
\tableline
\end{tabular}
\end{center}
\end{table}

The first order dispersion ($\Delta\lambda_0$) is clearly field-dependent:
it worsens along the diagonal from the W7 (the pointing with the highest dispersion)
to the W3 position (lowest dispersion) by 22$\%$ of the value at
position W1. Alternatively, it can be said that the amplitude of the field-dependence
between the centre of Chip 2 and the W3 and W7 corners is 11$\%$ of the value in the W1 
position.

\subsection{The Grism second order}

The grism second order spectra obtained in the W1 position and at the centre of
Chip 1 are plotted in Figure 4 for both WR~45 and WR~96.

\begin{figure}
\centerline{
\psfig{file=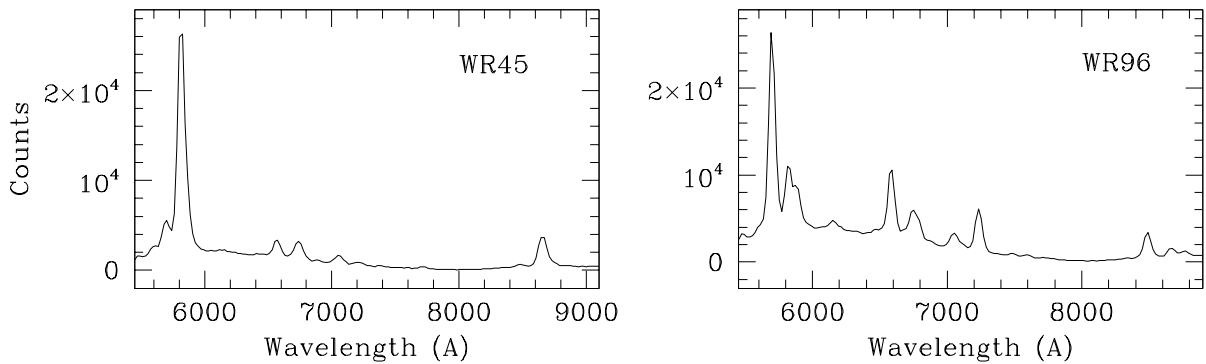}}
\caption{The grism second order spectra acquired for WR~45 at position W1 (left)
and WR~96 at the centre of Chip 1 (right).}
\end{figure}

The second order overlaps with the first order at $\lambda \simeq$ 5400 \AA\
and does not extend beyond 9000 \AA. For these reasons, the wavelength
solution was determined with a first-order polynomial fit, i.e.
$\lambda$ = $\lambda_0 +$ $\Delta\lambda_0$X, where X is again 
the distance along the dispersion axis from the
target X position in the direct image.

The results are listed in Table 5. As for Table 4, these values are
averages among multiple measurements available at each pointing. Typical
standard deviations are 0.1 \AA/pix and 7 \AA\ on the dispersion and
zero point, respectively. The RMS values of the fits are about 3 \AA.

\begin{table}
\caption{The wavelength solutions obtained for the grism second order
as a function of positon across the f.o.v of the WFC.}
\begin{center}
\begin{tabular}{lcc}
Position & $\lambda_0$ & $\Delta\lambda_0$ \\
         & (\AA)         & (\AA/pix)           \\
\tableline
W1   & 2432.38 & 20.75 \\
W2   & 2400.40 & 19.63 \\ 
W3   & 2445.01 & 22.85 \\
W4   & 2411.01 & 21.89 \\
W5   & 2405.74 & 20.54 \\
W7   & 2411.48 & 18.33 \\
W8   & 2391.95 & 19.13 \\
W9   & 2418.70 & 20.72 \\
W10  & 2409.33 & 21.56 \\
Chip 1 center& 2406.73 & 19.98\\
\tableline
\end{tabular}
\end{center}
\end{table}

The field dependence noticed earlier for the grism first order
is also present in the dispersion of the second. The amplitude
of the dispersion variation from center to the W7 and W3 corners
is about 11$\%$ of the dispersion in the W1 position. Once again, 
the dispersion decreases along the diagonal from W7 to W3.

\subsection{The Grism negative first order}

The negative first order spectra of WR~45 and WR~96 are shown in Figure
5.

\begin{figure}
\centerline{
\psfig{file=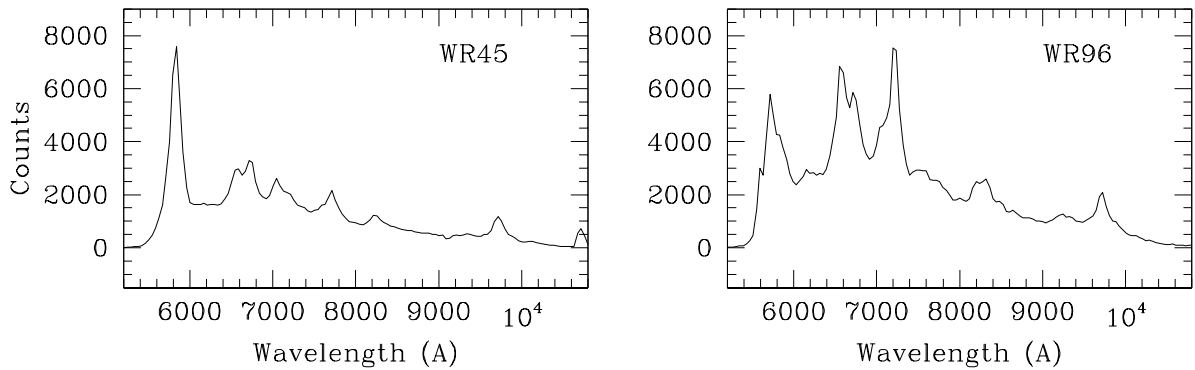}}
\caption{The grism negative first order spectra acquired for WR~45 at position W1 (left)
and WR~96 at the centre of Chip 1 (right).}
\end{figure}

Since the resolution is here lower than for the positive first
order and the noise higher, 
the wavelength solution of the negative first order was fitted
with a first-order polynomial, $\lambda$ = $\lambda_0 +$ $\Delta\lambda_0$X
where X is the distance in pixels along the dispersion axis from the target
X position in the direct image. The measurements of dispersion and zero
point obtained from multiple spectra acquired at the same pointing
were averaged and are presented in Table 6.

\begin{table}
\caption{The wavelength solutions obtained for the grism negative first order
as a function of positon across the f.o.v of the WFC.}
\begin{center}
\begin{tabular}{lcc}
Position & $\lambda_0$ & $\Delta\lambda_0$ \\
         & (\AA)         & (\AA/pix)           \\
\tableline
W1   & -4820.51 & -41.71 \\
W2   & -5026.08 & -40.05 \\ 
W3   & -4882.90 & -46.48 \\
W4   & -4808.92 & -44.37 \\
W5   & -4995.32 & -41.54 \\
W7   &  &  \\
W8   &  &  \\
W9   & -4858.67 & -41.96 \\
W10  & -4862.51 & -43.81 \\
Chip 1 center& -4784.49 & -40.14\\
\tableline
\end{tabular}
\end{center}
\end{table}  

Since W7 and W8 positions are closer to the edge of the field than
W5 (cf. Figure 2), the negative first order falls physically outside the frame.
Nevertheless, a variation in the dispersion of about 11$\%$ of the value in W1 
is still detected between the W1 and W3 positions.
The standard deviation is 0.1 \AA/pix and 27 \AA\ on the dispersion and
the zero point, respectively. The typical RMS of the first-order 
polynomial fits is 9 \AA.

\section{Products delivered to users}

The average dispersion coefficients derived for the ten positions across the
f.o.v of the WFC have to be parametrized as a function of position in order to
extract and calibrate spectra anywhere within the WFC aperture. We
thus derived a two dimensional fit for each parameter of the dispersion solutions 
of each grism order, where each parameter is a function of the (X,Y) coordinates
of the target in the direct image. These 2D fits were perfomed by adopting surface fits
polynomials. The same was also done for the X- and Y-shifts, the tilt of the spectra,
the orders separation and length.

The above fits are stored in calibration files used by the ST-ECF slitless
spectra extraction code aXe (Pirzkal et al. 2001a) and are delivered together
with the software  package.

Once the wavelength solution was determined, the flat-field
correction and the flux calibration using the SMOV and INTERIM spectra of
two White Dwarfs, GD~153 and G191B2B could be formalised. This is fully described 
in Pirzkal et al., this volume. It is also possible, at this stage of the calibrations, to
correct the extracted spectra for CCD fringing. The modeling of the WFC fringing
is explained in detail in Walsh et al., this volume.



\begin{references}
\reference van der Hucht, K.A., 2001, ``The VII$^{th}$ Catalogue of Galactic
Wolf-Rayet stars'', New AR, 45, 135 
\reference Pasquali, A., Pirzkal, N., Walsh, J.R., 2001a, ST-ECF ISR ACS 2001-04,
``Selection of Wavelength Calibration Targets for the ACS Grism'' 
\reference Pasquali, A., Pirzkal, N., Walsh, J.R., 2003, ST-ECF ISR ACS 2003-01,
``The in-orbit Wavelength Calibration of the WFC Grism'', in preparation 
\reference Pasquali, A., Pirzkal, N., Walsh, J.R., Hook, R.N., Freudling, W.,
Albrecht, R., Fosbury, R.A.E., 2001b, ST-ECF ISR ACS 2001-02, ``The Effective
Spectral Resolution of the WFC and HRC Grism''  
\reference Pirzkal, N., Pasquali, A., Demleitner, M., 2001a, ST-ECF Newsletter,
29, 5 
\reference Pirzkal, N., Pasquali, A., Walsh, J.R., Hook, R.N., Freudling, W.,
Albrecht, R., Fosbury, R.A.E., 2001b, ST-ECF ISR ACS 2001-01, ``ACS Grism
Simulations using SLIM 1.0''
\end{references}
\end{document}